\documentclass[aps,prb,reprint,superscriptaddress,amsmath,amssymb]{revtex4-2}

\usepackage{graphicx}    
\usepackage{dcolumn}    
\usepackage{bm}               
\usepackage[version=4]{mhchem} 
\usepackage{siunitx}
\usepackage{xcolor}
\usepackage{hyperref}
\usepackage{xr-hyper}
\usepackage{float}

\hypersetup{
    colorlinks=true,
    linkcolor=blue,
    filecolor=gray,      
    urlcolor=blue,
    citecolor=blue,
}
\begin{document}
\raggedbottom

\title{Frustrated Magnetism from Dimerized $S = 1/2$ Co$^{2+}$ in Rose-Phase Na$_2$CoP$_2$O$_7$}

\author{Shukun Li}
\affiliation{School of Physics and Astronomy, University of Minnesota, Twin Cities, Minneapolis, MN 55455, USA}

\author{Leon Ma}
\affiliation{School of Physics and Astronomy, University of Minnesota, Twin Cities, Minneapolis, MN 55455, USA}

\author{Xianghan Xu}
\email{xu001395@umn.edu} 
\affiliation{School of Physics and Astronomy, University of Minnesota, Twin Cities, Minneapolis, MN 55455, USA}

\date{\today}

\begin{abstract}
We report the growth of high-quality single crystals and the magnetic, thermodynamic, and dielectric properties of rose-phase Na$_2$CoP$_2$O$_7$, a low-symmetry magnet in which corner-sharing Co$^{2+}$ dimers form an approximately triangular network. Magnetic measurements reveal dominant antiferromagnetic interactions and pronounced anisotropy among effective $S=1/2$ Co$^{2+}$ moments. Broad low-temperature maxima in the magnetic susceptibility and heat capacity suggest the development of short-range magnetic correlations below approximately 4 K. The large separation between the absolute Curie--Weiss temperatures and the characteristic correlation temperature indicates strong magnetic frustration arising from competing interactions within the coupled-dimer network. Applied magnetic fields suppress these anomalies and reduce the recovered magnetic entropy, suggesting further enhancement of frustration under field. A systematic magnetodielectric response closely follows the field dependence of the magnetization, indicating coupling between magnetic and dielectric degrees of freedom. The low lattice symmetry enables multiple inequivalent exchange pathways and anisotropic interactions while relaxing symmetry constraints on coupling among spin, lattice, and dielectric degrees of freedom. These results establish rose-phase Na$_2$CoP$_2$O$_7$ as a platform for exploring frustrated quantum magnetism and coupled functionalities in effective $S=1/2$ magnets.
\end{abstract}

\maketitle 

\begin{figure*}[t]
\centering
\includegraphics[
        width=\textwidth,
        trim=0cm 2.5cm 0cm 2cm,
        clip]{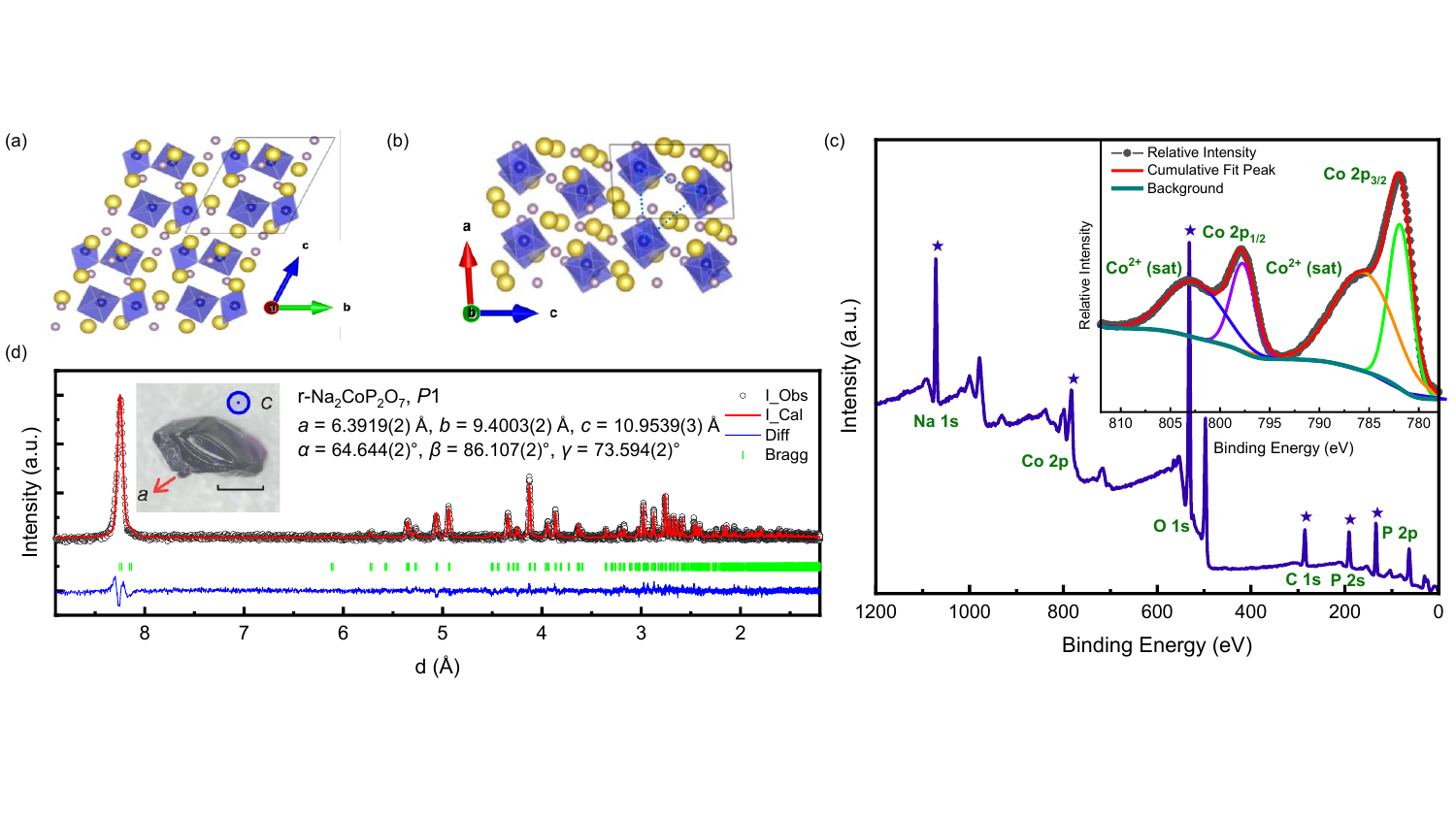}
\caption{
Structural and chemical characterization of r-Na$_2$CoP$_2$O$_7$.
(a) Crystal structure of r-Na$_2$CoP$_2$O$_7$
(space group $P1$) viewed along the crystallographic
$a$ axis. Yellow and purple spheres represent Na and P atoms, respectively; dark-blue spheres represent Co atoms, and the translucent blue polyhedra denote CoO$_6$ octahedra.
(b) Crystal structure viewed along the crystallographic
$b$ axis.
(c) X-ray photoelectron spectroscopy (XPS) survey spectrum.
The inset shows the high-resolution Co 2$p$ spectrum,
confirming the predominant Co$^{2+}$ oxidation state.
(d) Room-temperature powder x-ray diffraction pattern and
Rietveld refinement using the triclinic $P1$ structural model.
Open circles, red line, blue line, and green tick marks
represent the observed profile, calculated profile,
difference curve, and Bragg reflection positions,
respectively.
The inset shows a representative r-Na$_2$CoP$_2$O$_7$ single crystal used for Laue diffraction orientation; The red arrow and blue symbol indicate the crystallographic a and c directions, respectively. The scale bar represents 1 mm.
}
\label{fig:structure_xrd}
\end{figure*}

\section{Introduction}
\label{sec:intro}
Magnetic frustration provides a compelling setting for emergent quantum phenomena and unconventional collective excitations. Frustrated magnets have attracted extensive attention not only because they host exotic quasiparticles and novel magnetic states, but also because of their potential connections to broader topics ranging from high-temperature superconductivity to quantum information science \cite{Lacroix2011,Ramirez1994}. A major route toward magnetic frustration arises from lattice geometry. In antiferromagnetic systems built from, for example, triangular, kagome, or pyrochlore lattices, competing interactions often prevent all pairwise exchange interactions from being simultaneously satisfied, leading to highly degenerate ground states and enhanced quantum fluctuations \cite{Starykh2015,Greedan2001,Chamorro2021}. Beyond lattice geometry, spin-orbit coupling (SOC) provides another important ingredient for frustrated magnetism. By imposing strong directional constraints on spin degrees of freedom, SOC can effectively generate Ising-like anisotropy and further enhance magnetic frustration \cite{Balents2010}. Rare-earth magnets have therefore become a central focus of frustrated magnetism because their localized 4f electrons often exhibit strong SOC \cite{Harris1997,Morris2009,Gao2019,Sibille2015,Gao2026}. However, the highly localized nature of 4f orbitals typically results in relatively weak exchange interactions, pushing the relevant energy scales to very low temperatures and making experimental investigations of their magnetic excitations challenging.

In contrast, recent studies have demonstrated that octahedrally coordinated Co$^{2+}$ ions can provide an attractive alternative platform. The interplay between SOC and crystal electric field (CEF) splitting frequently produces an effective spin-$1/2$ ground state while maintaining substantially stronger exchange interactions than those found in most rare-earth systems. As a result, Co$^{2+}$-based magnets have emerged as a rich arena for quantum magnetism, exhibiting a variety of unconventional magnetic states and excitations on triangular, honeycomb, and related frustrated lattices \cite{Zhang2023BaCo2AsO4,Halloran2023,Zhong2020Honeycomb,Lin2021,Songvilay2020,Chen2026Supersolid,Xu2023Large}. An additional degree of freedom arises when magnetic ions form dimers. Dimerization can introduce competing intra- and interdimer interactions, enrich the magnetic phase diagram, and introduce an additional source of frustration beyond lattice geometry. Combined with the relatively large interaction energy scales characteristic of 3$d$ transition-metal systems, dimerized Co$^{2+}$ materials offer promising opportunities for both the experimental detection and potential applications of frustration-driven phenomena \cite{Zhong2020,Xu2024,Zhang2026}.

Here we investigate rose-phase Na$_2$CoP$_2$O$_7$ (r-Na$_2$CoP$_2$O$_7$), a dimerized cobalt pyrophosphate in which corner-sharing Co$^{2+}$ dimers are arranged on an approximately triangular network. High-quality single crystals were grown from molten, enabling a comprehensive study of its thermodynamic and functional properties. Anisotropic magnetization and heat-capacity measurements reveal dominant antiferromagnetic short-range correlations among effective $S=1/2$ Co$^{2+}$ moments and establish r-Na$_2$CoP$_2$O$_7$ as a strongly frustrated magnet. Dielectric measurements further reveal a pronounced anisotropic magnetodielectric response, dominated by the off-diagonal configuration with the electric and magnetic fields applied along different crystallographic directions. In addition, the crystal adopts the exceptionally low-symmetry space group $P1$, resulting in strongly distorted CoO$_6$ octahedra and a low-symmetry crystal-field environment that further modifies the spin-orbit-entangled states of Co$^{2+}$. Our results establish r-Na$_2$CoP$_2$O$_7$ as a rare example of a strongly interacting and highly frustrated Co$^{2+}$ magnet with pronounced magnetodielectric coupling, providing an attractive platform for exploring the interplay among short-range spin correlations, spin-orbit coupling, and dielectric degrees of freedom.

\section{Experimental Methods}
\label{sec:methods}
\subsection{Material Synthesis}
Na$_2$CoP$_2$O$_7$ is known to exhibit multiple polymorphs, including the low-temperature blue phase and the high-temperature rose phase \cite{Avdeev2019}. In our synthesis, crystallization directly from the melt yields the rose phase, which remains stable upon cooling to room temperature. Single crystals of r-Na$_2$CoP$_2$O$_7$ were grown by a slow cooling method from stoichiometric mixtures of Na$_2$CO$_3$, CoCO$_3$, and NH$_4$H$_2$PO$_4$ powders (Alfa Aesar, 99.99\%). The mixture was heated in a box furnace at a rate of \SI{180}{\celsius\per\hour} to \SI{400}{\celsius} and calcined in air for \SI{12}{\hour}. To prevent chemical reactions between the sample and the container at high temperatures, the product was reground then transferred into a Pt crucible, melted at \SI{750}{\celsius} at a rate of \SI{180}{\celsius\per\hour}, kept for 10 h, slowly cooled to \SI{600}{\celsius} at a rate of \SI{2}{\celsius\per\hour}, and naturally cooled to room temperature. The millimeter-size single crystals were mechanically separated from the product. The orientation is determined using a Photonic Science x-ray Laue diffractometer as shown in Fig.~S1 of the Supplemental Material~ \cite{SM} .
\subsection{X-ray Diffraction}
Powder x-ray diffraction measurements were performed at room temperature using a Rigaku SmartLab XE diffractometer with Cu K$\alpha$ radiation ($\lambda = 1.5406$ \AA). Diffraction data were collected over the $2\theta$ range of $10^\circ$--$80^\circ$. Rietveld refinements were carried out using the FullProf suite \cite{rodriguez1993recent}. Crystal structures were visualized using VESTA \cite{momma2011vesta}. Laue diffraction was performed using a Photonic Science diffractometer equipped with a reflection detector.

\subsection{Physical Property Measurements}
Magnetic measurements were performed using a Vibrating Sample Magnetometry (VSM) option in the Quantum Design Physical Property Measurement System (PPMS) on oriented crystals. The temperature dependence of the magnetic susceptibility $\chi(T)$ was measured between 2 and 300 K under an applied magnetic field of 0.1 T, and the Curie--Weiss analysis was performed using the 100 to 300 K range. Low-temperature magnetic susceptibility measurements were further carried out between 2 and 20 K under magnetic fields ranging from 0.1 to 9 T for $H \parallel a$, $H \perp  a$, $c$, and $H \parallel c$. Isothermal magnetization $M(H)$ measurements were performed at 2 K in magnetic fields from $-9$ to 9 T. 

Heat-capacity measurements were also performed using PPMS employing the two-tau method. Zero-field data were collected between 2 and 200 K. Additional low-temperature measurements were carried out between 2 and 10 K under magnetic fields ranging from 0 to 4.5 T applied perpendicular to the crystal rod axis, i.e. $H \parallel c$. The lattice contribution was estimated using a modified Debye model and subtracted from the total heat capacity to obtain the magnetic contribution. The magnetic entropy was determined by integrating the magnetic heat capacity divided by temperature.

Dielectric measurements were performed using a customized sample probe of PPMS and a Quadtech7600 LCR meter. Gold wires and silver epoxy were attached to opposite sides of the crystal. An AC excitation voltage of 1 V at a frequency of 44 kHz was applied during the measurements. The magnetic-field dependence of the dielectric response was measured under three field configurations: (i) $E\parallel a$ with $H$ perpendicular to the $a$ and $c$, (ii) $E\parallel c$ with $H\parallel a$, and (iii) $E\parallel H\parallel a$. According to Laue diffraction measurements, the crystal growth direction is approximately aligned with the crystallographic $a$ axis. The magnetodielectric response is presented as the normalized change in the dielectric constant, $\Delta\varepsilon(H)/\varepsilon(0)$.

\begin{figure*}[t]
\centering
\includegraphics[
        width=\textwidth,
        trim=0cm 0cm 0.5cm 0cm,
        clip]{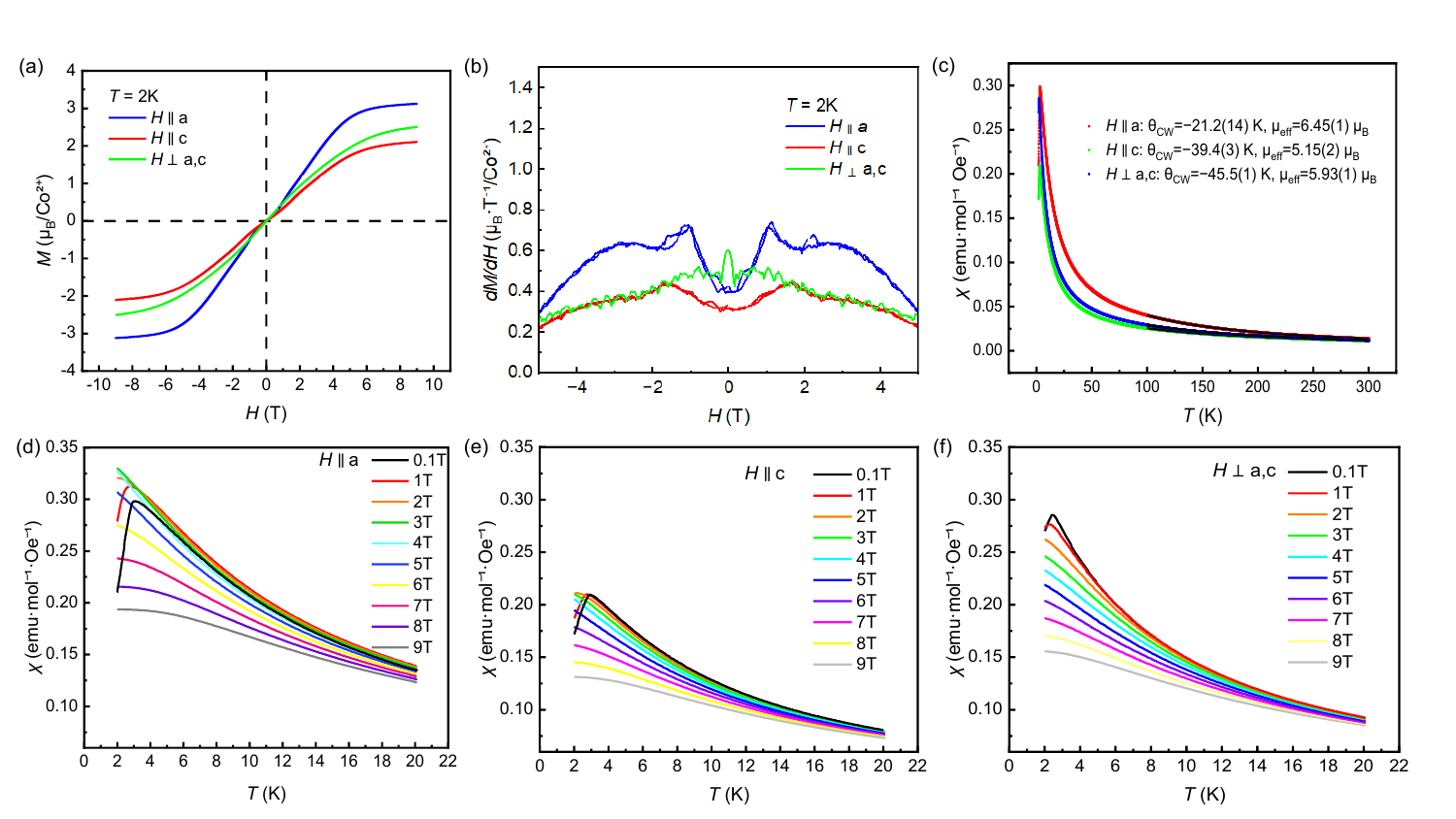}
\caption{
Magnetic properties of r-Na$_2$CoP$_2$O$_7$ for magnetic fields applied along three crystallographic directions. 
(a) Field dependence of the magnetization $M(H)$ measured at 2 K for $H\parallel a$, $H\parallel c$, and $H\perp a,c$.
(b) Differential susceptibility $dM/dH$ obtained from the numerical derivatives of the magnetization curves shown in (a).
(c) Temperature dependence of the magnetic susceptibility $\chi(T)$ measured under an applied field of 0.1 T for the three field orientations. The black solid lines represent the Curie--Weiss fits, and the corresponding Curie--Weiss temperatures $\theta_{\mathrm{CW}}$ and effective magnetic moments $\mu_{\mathrm{eff}}$ are indicated in the panel.
(d--f) Low-temperature magnetic susceptibility $\chi(T)$ measured under magnetic fields ranging from 0.1 to 9 T for (d) $H\parallel a$, (e) $H\parallel c$, and (f) $H\perp a,c$.
}
\label{fig:magnetic}
\end{figure*}

\section{Results and Discussion}
\label{sec:results}

\subsection{Crystal Structure and Characterization}

Figure~\ref{fig:structure_xrd} summarizes the crystal structure, powder x-ray diffraction results, and chemical characterization of the r-Na$_2$CoP$_2$O$_7$ crystal. Fig. 1(d) shows that the observed diffraction pattern can be successfully refined using the previously reported triclinic $P1$ structural model~ \cite{Avdeev2019}. The refined lattice parameters are $a = 6.3919(2)$~\AA, $b =9.4003(2)$~\AA, $c = 10.9539(3)$~\AA, $\alpha = 64.644(2)^\circ$, $\beta = 86.107(2)^\circ$, and $\gamma = 73.594(2)^\circ$. The refined fractional atomic coordinates and site occupancies are summarized in Table~S1 of the Supplemental Material.

The crystal structure is shown in Fig.~\ref{fig:structure_xrd}(a,b). The magnetic Co$^{2+}$ ions are octahedrally coordinated by oxygen, forming CoO$_6$ octahedra. Because the crystal adopts the $P1$ space group, no symmetry constraints are imposed on the local coordination, resulting in highly distorted CoO$_6$ octahedra. These octahedra form corner-sharing dimers along the $b$ axis, while the dimers are arranged in a nearly triangular lattice in the plane perpendicular to $b$. This crystal structure incorporates several key ingredients for complex magnetism, including low local symmetry, magnetic dimerization, and geometric frustration associated with the nearly triangular network.

The chemical composition and valence state of r-Na$_2$CoP$_2$O$_7$ were further examined by x-ray photoelectron spectroscopy, as shown in Fig.~\ref{fig:structure_xrd}(c). The survey spectrum confirms the presence of Na, Co, P, and O, consistent with the nominal chemical composition of r-Na$_2$CoP$_2$O$_7$. The high-resolution Co 2$p$ spectrum is shown in the inset of Fig.~\ref{fig:structure_xrd}(c). The Co 2$p_{3/2}$ and Co 2$p_{1/2}$ peaks are accompanied by pronounced satellite features characteristic of Co$^{2+}$. These features indicate that cobalt is predominantly present in the divalent state, consistent with previous XPS studies of triclinic Na$_2$CoP$_2$O$_7$. \cite{Gond2022} The determined 2+ valence of Co ions is consistent with the magnetic behavior discussed below.

\subsection{Magnetic Properties}

The magnetic properties of r-Na$_2$CoP$_2$O$_7$ along three special orientations are summarized in Fig.~\ref{fig:magnetic}, where $H \parallel a$, $H \perp  a$, $c$, and $H \parallel c$. This basis is chosen because the $a$ and $c$ axes are nearly perpendicular to each other in this unit cell. The field dependence of the magnetization measured at 2 K is shown in Fig.~\ref{fig:magnetic}(a). For all three field orientations, the magnetization increases continuously with increasing magnetic field and approaches saturation near 9 T to 2 - 3 $mu_{\mathrm{B}}$, close to the expected spin-only saturation moment. Noticeable changes in slope are observed in the low-field magnetization curves for $H\parallel a$ and $H \parallel c$, whereas the response for $H \perp  a$, $c$ is comparatively smooth. This contrast is more clearly revealed in the corresponding differential susceptibility $dM/dH$ shown in Fig.~\ref{fig:magnetic}(b), where pronounced symmetric low-field anomalies are observed for $H\parallel a$ and $H \parallel c$, while no comparable anomaly is evident for $H \perp  a$, $c$. The temperature dependence of the magnetic susceptibility is shown in Fig.~\ref{fig:magnetic}(d--f). For all three field orientations, the susceptibility increases upon cooling and exhibits a broad maximum at low temperatures under relatively weak magnetic fields. Such behavior is associated with antiferromagnetic short range correlations, which is further supported by the heat capacity data in the next section. With increasing magnetic field, the broad maximum is progressively suppressed and shifts toward lower temperatures or outside the measured temperature range, indicating that the low-temperature magnetic correlations are strongly field dependent. Compared with the responses for other two directions, the susceptibility for $H \perp  a$, $c$ has a less pronounced anomaly and evolves more smoothly with increasing field.

To further characterize the magnetic interactions, the susceptibilities measured under an applied field of 0.1 T were analyzed using the Curie--Weiss law, as shown in Fig.~\ref{fig:magnetic}(c). The resulting Curie--Weiss temperatures are $\theta_{\mathrm{CW}}=-21.22(14)$ K, $-39.4(3)$ K, and $-45.5(1)$ K for $H\parallel a$, $H \parallel c$, and $H \perp  a$, $c$, respectively. The negative values of $\theta_{\mathrm{CW}}$ indicate that antiferromagnetic interactions are dominant for all three field orientations. The corresponding effective magnetic moments are $\mu_{\mathrm{eff}}=6.45(1)\mu_{\mathrm{B}}$, $5.15(2)\mu_{\mathrm{B}}$, and $5.93(1)\mu_{\mathrm{B}}$, respectively. These values are larger than the value expected for high-spin Co$^{2+}$. Such an enhancement may arise from a substantial orbital contribution associated with spin--orbit coupling, and the detectable orientation dependence of both $\theta_{\mathrm{CW}}$ and $\mu_{\mathrm{eff}}$ indicates an anisotropic $g$ tensor of  Co$^{2+}$ in r-Na$_2$CoP$_2$O$_7$.

\subsection{Heat-Capacity Properties}

\begin{figure*}[t]
\centering
\includegraphics[
        width=\textwidth,
        trim=4.5cm 0.5cm 4.5cm 0.5cm,
        clip]{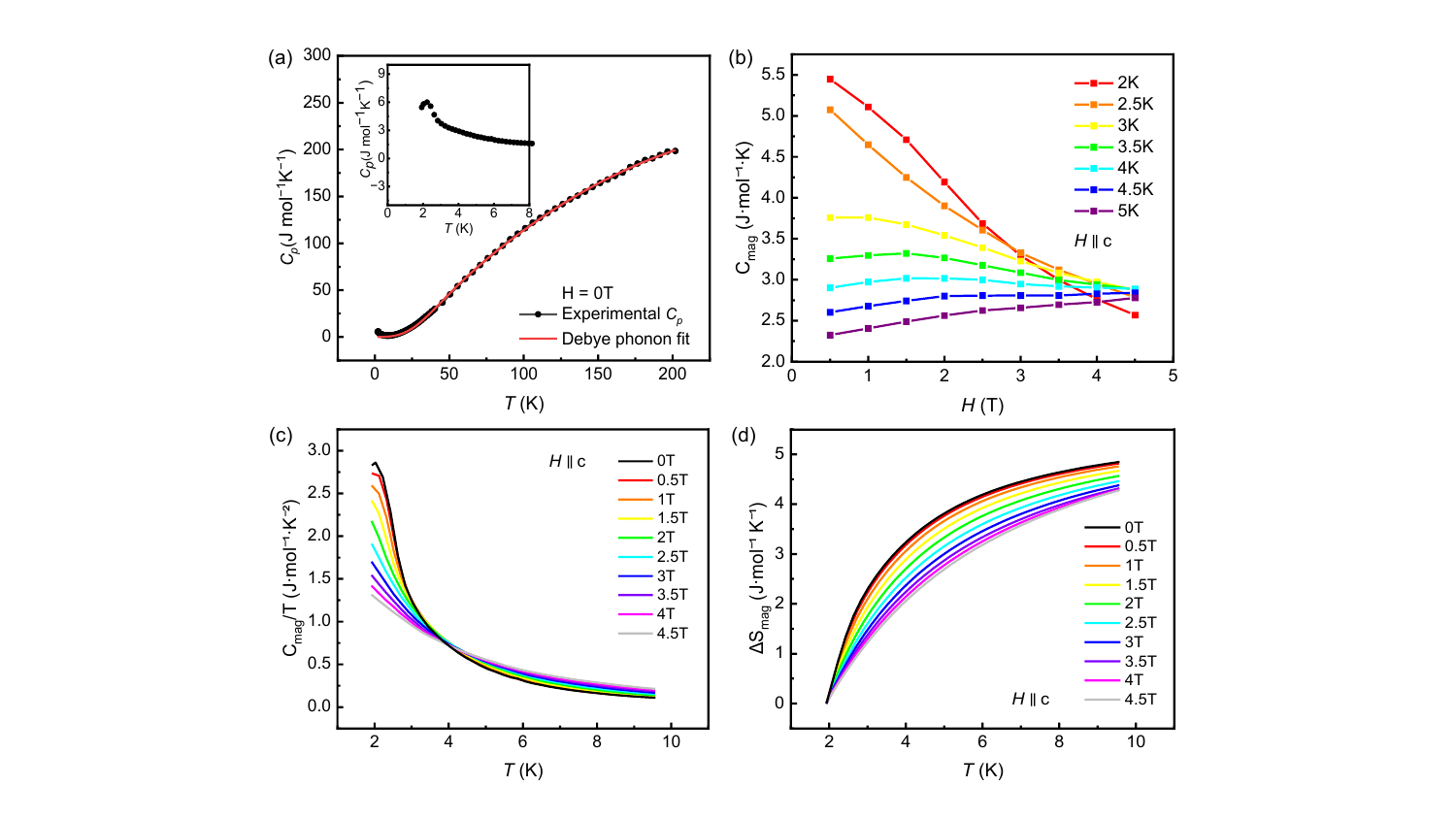}
\caption{
Heat-capacity properties of r-Na$_2$CoP$_2$O$_7$ measured with the magnetic field applied perpendicular to the crystal growth direction ($H\parallel c$) 
(a) Temperature dependence of the heat capacity $C_p$ measured at zero magnetic field together with the Debye-phonon fit. The inset shows $C_p/T$ below 8 K.
(b) Magnetic heat capacity $C_{\rm mag}$ plotted as a function of magnetic field at selected temperatures between 2 and 5 K, reconstructed from heat-capacity measurements performed under different magnetic fields. 
(c) Temperature dependence of $C_{\rm mag}/T$ measured under magnetic fields ranging from 0 to 4.5 T. 
(d) Magnetic entropy $\Delta S_{\rm mag}$ obtained from the integration of $C_{\rm mag}/T$ under different magnetic fields.
}
\label{fig:heatcapacity}
\end{figure*}

Heat capacity measurements on r-Na$_2$CoP$_2$O$_7$ are performed to study the magnetic correlations and magnetic entropy. Figure~\ref{fig:heatcapacity}(a) shows the zero-field heat capacity data from 2 K to 200 K. The high-temperature data between 30 and 200 K can be well fitted by a two-component Debye phonon model,

\begin{equation}
C_{\mathrm{lat}}
=
9R\sum_{i=1}^{2}
N_i
\left(\frac{T}{\Theta_{Di}}\right)^3
\int_0^{\Theta_{Di}/T}
\frac{x^4 e^x}{(e^x-1)^2}\,dx ,
\end{equation}

where $N_i$ and $\Theta_{Di}$ denote the effective number of atoms and the Debye temperature associated with the $i$th phonon contribution, respectively. The lower-Debye-temperature component predominantly captures lower-frequency lattice vibrations involving the heavier Na and Co atoms and softer collective modes, whereas the higher-Debye-temperature component primarily accounts for higher-frequency vibrations associated with the lighter atoms and the relatively stiff P--O framework. The fit yields characteristic Debye temperatures of $\Theta_{D1}=251.13$~K and $\Theta_{D2}=731.45$~K, with corresponding effective oscillator weights of $N_1=4.60$ and $N_2=6.67$, respectively. The sum $N_1+N_2=11.27$ is reasonably close to the 12 atoms per formula unit of Na$_2$CoP$_2$O$_7$. The low-temperature heat capacity deviates from the phonon-only contribution, suggesting the onset of magnetic correlations.Although the zero-field heat capacity exhibits a maximum near 2.2 K, the anomaly is broad rather than the sharp $\lambda$-type feature typically associated with a conventional long-range magnetic ordering transition. This broad feature therefore more likely reflects the development of short-range magnetic correlations rather than the onset of long-range magnetic order. The field dependence of the heat capacity was further investigated with the magnetic field applied along the crystallographic \(c\) axis ($H\parallel c$). The $M$--$H$ measurements show field-induced anomalies for both this field orientation and $H \parallel a$. However, because the crystal rods grow roughly along the $a$ axis, mounting the sample for $H \parallel a$ is difficult. Therefore, the $H\parallel c$ configuration was chosen. After subtracting the phonon contribution, the magnetic heat capacity $C_{\rm mag}$ was extracted. Figure~\ref{fig:heatcapacity}(b) shows the field dependence of the heat capacity at different temperatures. The applied magnetic field increases the magnetic heat capacity above 4~K, whereas it reduces the magnetic entropy below 4~K. This crossover temperature coincides with the temperature at which the magnetic susceptibility reaches its maximum. The temperature dependence of $C_{\rm mag}$ in Fig.~\ref{fig:heatcapacity}(c) exhibits the same trend, and at low temperature the magnetic correlations are gradually suppressed with increasing magnetic field. The magnetic entropy $\Delta S_{\rm mag}$ obtained from the integration of $C_{\rm mag}/T$ is shown in Fig.~\ref{fig:heatcapacity}(d). At zero field, the recovered magnetic entropy reaches 4.848 J mol$^{-1}$ K$^{-1}$ at 9.57 K, corresponding to approximately 84\% of $R\ln2$. Since the entropy integration was performed from the lowest measured temperature of approximately 2 K without extrapolation to 0 K, this value represents the entropy recovered between 2 and 9.57 K and therefore provides a lower bound for the total magnetic entropy. The recovered entropy approaching $R\ln2$ suggests that the low-temperature magnetic degrees of freedom are dominated by an effective doublet, consistent with an effective spin-$\frac{1}{2}$ ground state of the Co$^{2+}$ ions.

\begin{figure*}[t]
    \centering
    \includegraphics[
        width=\textwidth,
        trim=3cm 3cm 3cm 4cm,
        clip
    ]{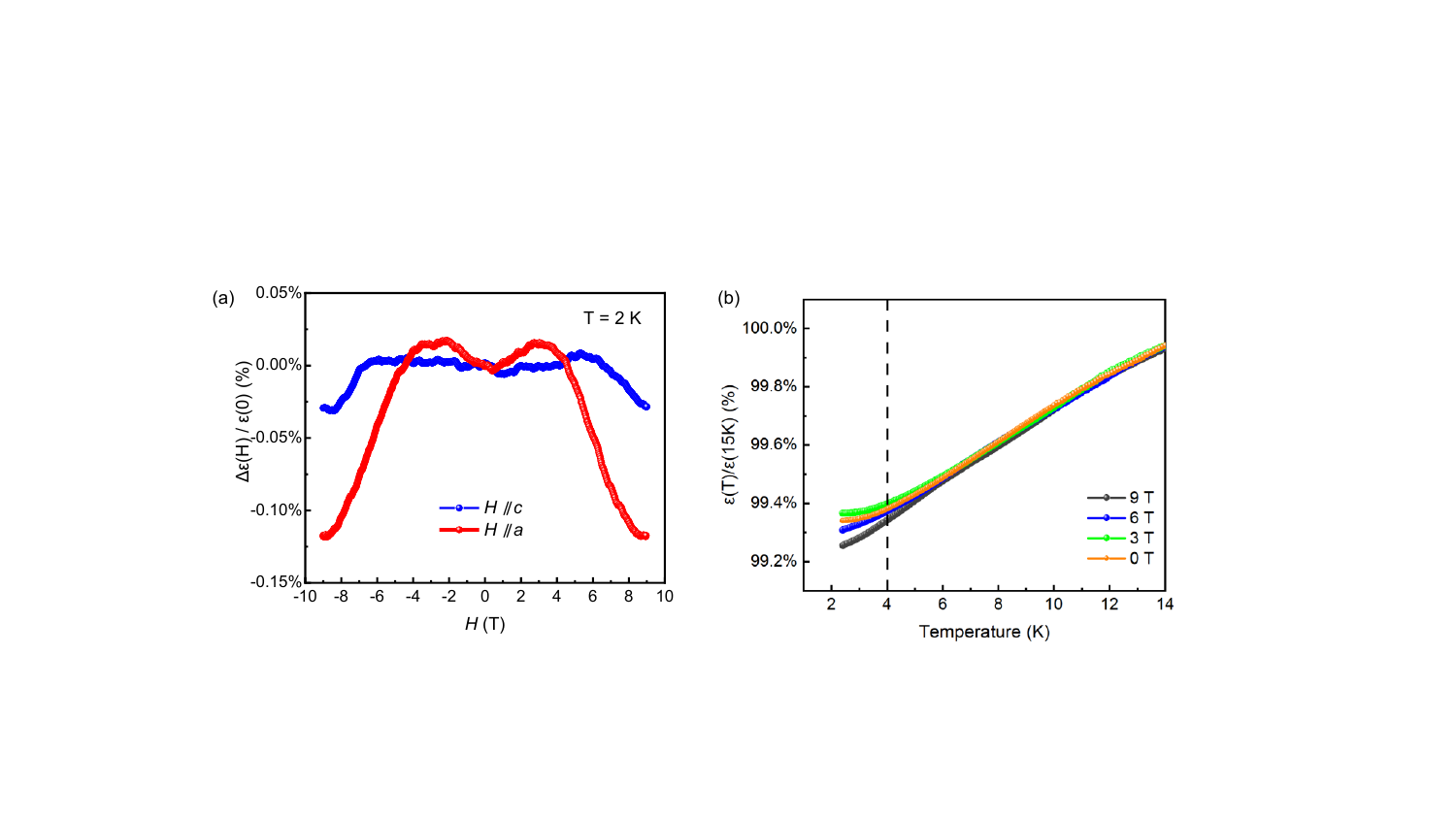}
    
    \caption{
    Dielectric response of Na$_2$CoP$_2$O$_7$ for $E \parallel c$.
(a) Magnetic-field dependence of the relative dielectric change,
$\Delta\varepsilon(H)/\varepsilon(0)$, measured at 2 K for $H \parallel a$ and $H \parallel c$.
(b) Temperature dependence of the normalized dielectric constant,
$\varepsilon(T)/\varepsilon(15\,\mathrm{K})$, measured at 0, 3, 6, and 9 T for $H \parallel a$.
The vertical dashed line in (b) indicates the broad maximum near 4 K in the susceptibility curves.
    }
    
    \label{fig:dielectric}
\end{figure*}

\subsection{Dielectric Properties}

The low crystallographic symmetry of r-Na$_2$CoP$_2$O$_7$ may allow coupling between its magnetic and dielectric degrees of freedom. Figure~\ref{fig:dielectric}(a) shows the magnetic-field dependence of the relative dielectric change,
$\Delta\varepsilon(H)/\varepsilon(0)=[\varepsilon(H)-\varepsilon(0)]/\varepsilon(0)$,
measured with the electric field applied along the crystallographic $c$ axis and the magnetic field applied along the $a$ axis. Overall, the response for a field applied along the $a$ axis is much stronger than that for a field along the $c$ axis, suggesting a dominant off-diagonal magnetodielectric response. In the off-diagonal channel, the relative dielectric change initially increases with field, then decreases and becomes negative at higher fields, reaching approximately $-0.12\%$ near $\pm9$~T. The characteristic kink occurs at a field close to the anomaly observed in the M-H curve, suggesting that the magnetodielectric response is coupled to the field evolution of short-range magnetic correlations. The response is approximately symmetric with respect to field reversal, with no obvious hysteresis between increasing and decreasing field sweeps.

Figure~\ref{fig:dielectric}(b) presents the temperature dependence of the dielectric constant along the $c$ axis measured under magnetic fields of 0, 3, 6, and 9~T applied along the $a$ axis. The dielectric constant is normalized by its value at 15~K, $\varepsilon(T)/\varepsilon(15~\mathrm{K})$. For all applied fields, the normalized dielectric constant increases smoothly with increasing temperature over the measured range. The curves begin to deviate from each other below approximately 4~K, while no detectable field dependence is observed at higher temperatures. This onset temperature coincides with the development of short-range magnetic correlations inferred from the magnetic susceptibility and heat-capacity measurements, providing further evidence that the low-temperature magnetodielectric response originates from the coupling between the dielectric properties and short-range magnetic correlations. Magnetodielectric effects in the absence of long-range magnetic order have also been reported in several frustrated magnetic systems, such as Na$_5$Co$_{15.5}$Te$_6$O$_{36}$ \cite{Saha2023} and K$_2$Co$_2$(SO$_4$)$_3$ \cite{xu2025anomalous}. These observations suggest that long-range magnetic order is not a prerequisite for magnetodielectric coupling and that short-range spin correlations can provide a sufficient magnetic background for such a response. In $r$-Na$_2$CoP$_2$O$_7$, the strong spin-orbit coupling of Co$^{2+}$ may play an important role in coupling these magnetic correlations to the dielectric response. Additional measurements under $E\parallel a$ with $H$ perpendicular to both the $a$ and $c$ axes, and under $E\parallel H\parallel a$, are presented in Fig.~S3 of the Supplemental Material.

\subsection{Discussion and Conclusion}

Overall, the magnetic susceptibility exhibits a broad anomaly near 4 K, while the heat capacity shows a weak peak-like anomaly near 2.22 K. These results indicate the development of strong short-range magnetic correlations below approximately 4 K, with an additional low-temperature anomaly emerging near 2.22 K. A similar short-range-correlated magnetic ground state, characterized by comparable susceptibility and heat-capacity behavior, has been reported in $\gamma$-BaCo$_2$(PO$_4$)$_2$ \cite{Nair2018}. Considering the large absolute value of the Curie--Weiss temperature, the magnetic ground state is highly frustrated. The fact that the anomaly is readily suppressed by an applied magnetic field further indicates the antiferromagnetic nature of the short-range correlations. Interestingly, even in the absence of long-range magnetic order, the short-range correlations exhibit pronounced anisotropy. As shown in Figs.~2(d--f), the susceptibility anomaly is strongest for $H \parallel a$, indicating that the $a$ axis is the easy axis of the short-range antiferromagnetic correlations. This interpretation is further supported by the $M$--$H$ data in Figs.~2(a,b), where the field applied along the $a$ axis produces the most pronounced kink. We associate this anomaly with a characteristic field scale over which the short-range magnetic correlations are strongly modified or suppressed, consistent with the disappearance of both the susceptibility anomaly and the magnetic heat-capacity anomaly under magnetic fields of similar magnitude. Furthermore, the heat-capacity data in Fig.~3 demonstrate that the applied magnetic field suppresses the magnetic entropy, indicating an enhancement of magnetic frustration. Similar behavior has been reported in the field-induced quantum spin liquid candidates Na$_2$Co$_2$TeO$_6$ and K$_2$Co$_2$TeO$_6$ \cite{Yao2020,Xu2023K2Co2TeO6}. 

The negative Curie--Weiss temperatures indicate predominantly antiferromagnetic interactions between the Co$^{2+}$ moments. However, the characteristic temperature of the broad low-temperature susceptibility maximum is much smaller than $|\theta_{\rm CW}|$, suggesting that the development of long-range order is strongly suppressed by competing interactions. Structurally, the Co$^{2+}$ ions form corner-sharing dimers that are further connected into an approximately triangular network. The absence of a clear isolated-dimer-like magnetic response suggests that the interdimer coupling is appreciable and that the system is better viewed as a network of coupled magnetic dimers rather than weakly interacting isolated dimers. Antiferromagnetic interdimer interactions on the nearly triangular network then provide a natural microscopic origin for the pronounced short-range correlations and magnetic frustration. A related example is CaCoV$_2$O$_7$, where neighboring Co$^{2+}$ moments form ferromagnetically coupled dimers with sizable antiferromagnetic interdimer interactions \cite{Murasaki2021}. Broad maxima in both the magnetic susceptibility and heat capacity occur above $T_N$ and are associated with the development of short-range magnetic correlations, before long-range antiferromagnetic order eventually sets in at lower temperature. In comparison, the nearly triangular arrangement of coupled dimers in the present system may introduce stronger competition among the interdimer interactions, further suppressing long-range order in favor of short-range correlations.

In summary, we synthesized high-quality single crystals of rose-phase Na$_2$CoP$_2$O$_7$ and established it as a frustrated magnet based on effective $S=1/2$ Co$^{2+}$ moments. Magnetic and thermodynamic measurements reveal predominantly antiferromagnetic interactions and pronounced short-range correlations, together with a weak heat-capacity anomaly near 2.22 K whose microscopic origin remains to be clarified. These short-range magnetic correlations further induce a sizable magnetodielectric coupling, which may originate from the strong spin--orbit coupling of Co$^{2+}$. Our results establish rose-phase Na$_2$CoP$_2$O$_7$ as a distinct platform for exploring frustration and spin--lattice coupling in spin--orbit-coupled $S=1/2$ magnets.

\begin{acknowledgments}
S. Li and X. Xu were supported by the U.S. Department of Energy (DOE) through the University of Minnesota Center for Quantum Materials under Grant No. DE-SC-0016371, as well as by startup support from the University of Minnesota. The authors acknowledge the use of instrumentation at the University of Minnesota Characterization Facility (CharFac).
\end{acknowledgments}

\bibliographystyle{apsrev4-2}
\bibliography{references}
\end{document}


\renewcommand{\thefigure}{S\arabic{figure}}
\renewcommand{\thetable}{S\arabic{table}}

\title{Supplementary Material for ``Frustrated Magnetism from Dimerized $S = 1/2$ Co$^{2+}$ in Rose-Phase Na$_2$CoP$_2$O$_7$''}

\maketitle

\section{Laue Diffraction}

\begin{figure*}[t]
    \centering
    \includegraphics[width=0.90\textwidth]{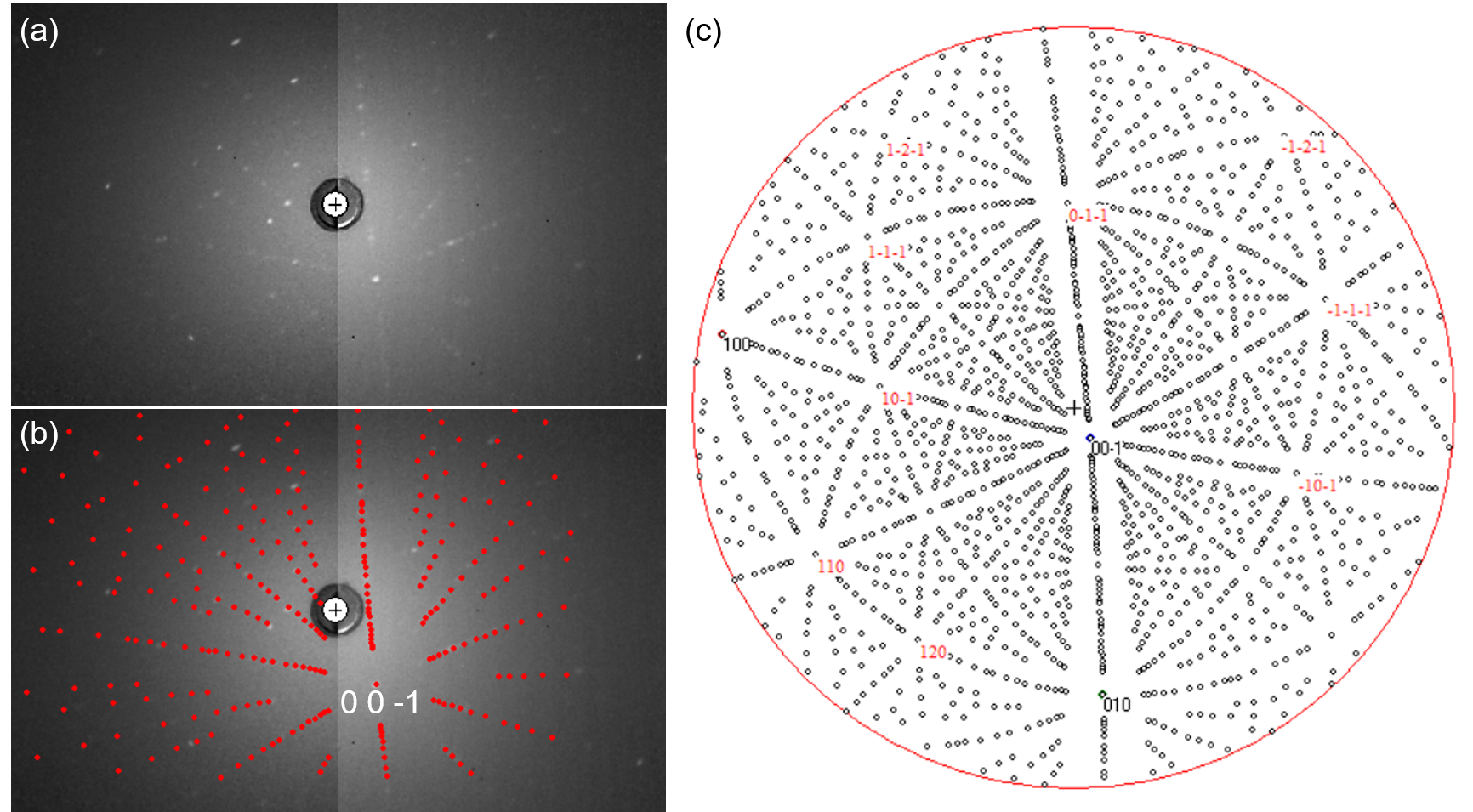}
    \caption{
    (a) Experimental back-reflection Laue diffraction pattern of an
    r-Na$_2$CoP$_2$O$_7$ single crystal.
    (b) Indexed Laue diffraction pattern with the calculated diffraction
    spots superimposed on the experimental image.
    (c) Simulated Laue diffraction pattern for the determined crystal
    orientation.
    }
    \label{fig:laue}
\end{figure*}

Figure~\ref{fig:laue} shows the experimental Laue diffraction pattern,
the indexed diffraction pattern, and the corresponding simulated pattern
for an r-Na$_2$CoP$_2$O$_7$ single crystal. The agreement between the
experimental and simulated patterns confirms the crystallographic
orientation of the crystal. The direction normal to the measured crystal
surface was indexed as approximately the crystallographic
$[00\bar{1}]$ direction. The sharp and well-defined diffraction spots
also indicate the good crystallinity of the measured single crystal.
This orientation analysis was used to align the sample for the subsequent
direction-dependent magnetic and dielectric measurements.

\section{Crystallographic Parameters}

The crystal structure of r-Na$_2$CoP$_2$O$_7$ was refined using powder
X-ray diffraction data in the triclinic space group $P1$.
The refined fractional atomic coordinates and site occupancies are listed
in Table~\ref{tab:atomic_coordinates}. 

\begin{table*}[t]
\centering
\caption{
Wyckoff positions, fractional atomic coordinates, and site occupancies
obtained from the Rietveld refinement of powder X-ray diffraction data
for r-Na$_2$CoP$_2$O$_7$. The structure was refined in the triclinic
space group $P1$ (No.~1), with lattice parameters
$a=6.3919(2)$~\AA,
$b=9.4003(2)$~\AA,
$c=10.9539(3)$~\AA,
$\alpha=64.644(2)^\circ$,
$\beta=86.107(2)^\circ$, and
$\gamma=73.594(2)^\circ$.
}
\label{tab:atomic_coordinates}

\renewcommand{\arraystretch}{1.2}
\setlength{\tabcolsep}{14pt}

\begin{tabular}{lccccc}
\toprule
Atom & Wyck. & $x/a$ & $y/b$ & $z/c$ & Occ. \\
\midrule
Na1 & $2i$ & 0.13742 & 0.77373 & 0.03086 & 1.000 \\
Na2 & $2i$ & 0.35189 & 0.14019 & 0.72385 & 1.000 \\
Co1 & $2i$ & 0.09241 & 0.60130 & 0.36845 & 1.000 \\
P1  & $2i$ & 0.25516 & 0.16188 & 0.39076 & 1.000 \\
P2  & $2i$ & 0.42792 & 0.33370 & 0.13968 & 1.000 \\
O1  & $2i$ & 0.06766 & 0.31113 & 0.42550 & 1.000 \\
O2  & $2i$ & 0.09741 & 0.90794 & 0.34181 & 1.000 \\
O3  & $2i$ & 0.25504 & 0.49549 & 0.17225 & 1.000 \\
O4  & $2i$ & 0.27343 & 0.54109 & 0.80217 & 1.000 \\
O5  & $2i$ & 0.34420 & 0.16139 & 0.23146 & 1.000 \\
O6  & $2i$ & 0.34825 & 0.16599 & 0.97304 & 1.000 \\
O7  & $2i$ & 0.49668 & 0.75792 & 0.48576 & 1.000 \\
\bottomrule
\end{tabular}
\end{table*}

\section{Additional Dielectric Measurements}

\begin{figure*}[t]
    \centering
    \includegraphics[width=0.90\textwidth]{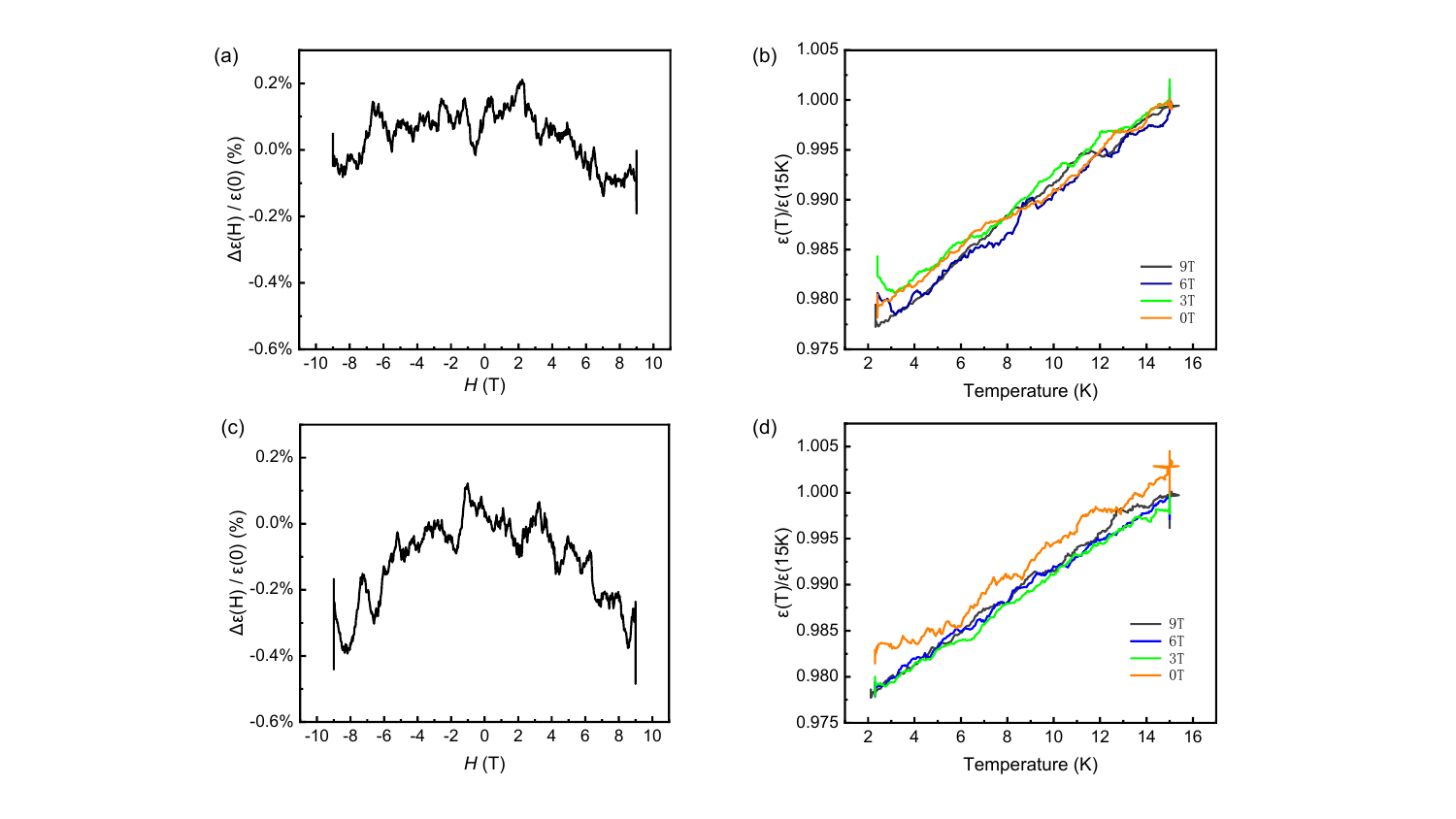}
    \caption{
Additional dielectric measurements of r-Na$_2$CoP$_2$O$_7$ for two
field configurations.
(a) Magnetic-field dependence of the relative dielectric change,
$\Delta\varepsilon(H)/\varepsilon(0)$, expressed as a percentage,
for $E\parallel a$ and $H\perp a$ and $c$.
(b) Temperature dependence of the normalized dielectric constant,
$\varepsilon(T)/\varepsilon(15~\mathrm{K})$, $E\parallel a$ and $H\perp a$ and $c$,
measured at 0, 3, 6, and 9~T.
(c) Magnetic-field dependence of
$\Delta\varepsilon(H)/\varepsilon(0)$, expressed as a percentage,
for $E\parallel H\parallel a$.
(d) Temperature dependence of
$\varepsilon(T)/\varepsilon(15~\mathrm{K})$ for the same configuration,
measured at 0, 3, 6, and 9~T.
All data were smoothed using a 20-point adjacent-averaging window.
}
    \label{fig:dielectric_supp}
\end{figure*}

Additional dielectric measurements were performed under two field
configurations to examine the anisotropy of the dielectric response of
r-Na$_2$CoP$_2$O$_7$. Figures~\ref{fig:dielectric_supp}(a) and
\ref{fig:dielectric_supp}(b) show the results for
$E\parallel a$ and $H\perp a,c$, while
Figs.~\ref{fig:dielectric_supp}(c) and
\ref{fig:dielectric_supp}(d) show the results for
$E\parallel H\parallel a$. All data were processed using a 20-point adjacent- average window.
